\documentclass[aps,prb,twocolumn,showpacs]{revtex4}
\usepackage{graphicx}

\begin{document}

\title{$I$-$V$ characteristics of the vortex state in MgB$_2$ thin films}

\author{Huan Yang,$^1$ Ying Jia,$^1$ Lei Shan,$^1$ Yingzi Zhang$^1$ and Hai-Hu
Wen$^1$}\email{hhwen@aphy.iphy.ac.cn} \affiliation{$^1$National
Laboratory for Superconductivity, Institute of Physics and National
Laboratory for Condensed Matter Physics, Chinese Academy of
Sciences, P.O.~Box 603, Beijing 100080, P.~R.~China}

\author{Chenggang Zhuang,$^{2,3}$ Zikui Liu,$^4$ Qi Li,$^2$ Yi Cui$^2$ and Xiaoxing Xi$^{2,4}$}
\affiliation{$^2$Department of Physics, The Pennsylvania State
University, University Park, Pennsylvania 16802, USA}

\affiliation{$^3$Department of Physics, Peking University, Beijing
100871, PR China}

\affiliation{$^4$Department of Materials Science and Engineering,
The Pennsylvania State University, University Park, Pennsylvania
16802, USA}

\date{\today}

\begin{abstract}
The current-voltage ($I$-$V$) characteristics of various MgB$_2$
films have been studied at different magnetic fields parallel to
$c$-axis. At fields $\mu_0H$ between 0 and 5$\;$T, vortex
liquid-glass transitions were found in the $I$-$V$ isotherms.
Consistently, the $I$-$V$ curves measured at different temperatures
show a scaling behavior in the framework of quasi-two-dimension
(quasi-2D) vortex glass theory. However, at $\mu _{0} H \ge 5\;$T, a
finite dissipation was observed down to the lowest temperature here,
$T=1.7\;$K, and the $I$-$V$ isotherms did not scale in terms of any
known scaling law, of any dimensionality. We suggest that this may
be caused by a mixture of $\sigma$ band vortices and $\pi$ band
quasiparticles. Interestingly, the $I$-$V$ curves at zero magnetic
field can still be scaled according to the quasi-2D vortex glass
formalism, indicating an equivalent effect of self-field due to
persistent current and applied magnetic field.
\end{abstract}

\maketitle

\section{Introduction}

Since the discovery of the two-gap superconductor MgB$_2$ in
2001,\cite{MgB2found} the mechanism of its superconductivity and
vortex dynamics has attracted considerable interests. The two
three-dimension (3D) $\pi$ bands and two quasi-two-dimension
(quasi-2D) $\sigma$ bands in this simple binary compound seem to
play an important role in the
superconductivity,\cite{coherence_length} as well as the normal
state properties.\cite{MR_normal,LiQ} The two sets of bands have
different energy gaps, i.e., about 7$\;$meV for the $\sigma$ bands,
and about 2$\;$meV for the $\pi$ bands.\cite{gaps1,gaps2} And the
coherent length of the $\pi$ bands is much larger than that of the
$\sigma$ bands \cite{coherence_length}. Many experiments have
demonstrated that the $\pi$-band superconductivity is induced from
the $\sigma$-band and there is a rich evidence for both the
interband and intraband scattering. Owing to the complicated nature
of superconductivity in this system, its vortex dynamics may exhibit
some interesting or novel features. Among various experimental
methods, measuring the current-voltage ($I$-$V$) characteristics at
different temperatures and magnetic fields can provide important
information for understanding the physics of the vortex state. Up to
now, the transport properties of MgB$_2$ have been studied on both
polycrystalline bulk samples\cite{MgB2IVpoly} and thin
films\cite{MgB2IVfilm}. In both cases, the $I$-$V$ characteristics
demonstrated good agreement with the 3D vortex glass (VG) theory.
This was partially due to the limited magnetic fields in the
experiment. In addition, it has been shown that the properties of
MgB$_2$ are very sensitive to the impurities and defects introduced
in the process of sample preparation, and the vortex dynamics must
be influenced, too. Therefore, it is necessary to investigate the
vortex dynamics in high quality MgB$_2$ epitaxial thin films and to
reveal the intrinsic properties of the vortex matter in this
interesting multiband system. In this paper, we present the $I$-$V$
characteristics of high-quality MgB$_2$ thin films measured at
various temperatures and magnetic fields. The vortex dynamics in
this system is then investigated in detail.

\section{Experiment}

\begin{figure}
\includegraphics[width=8cm]{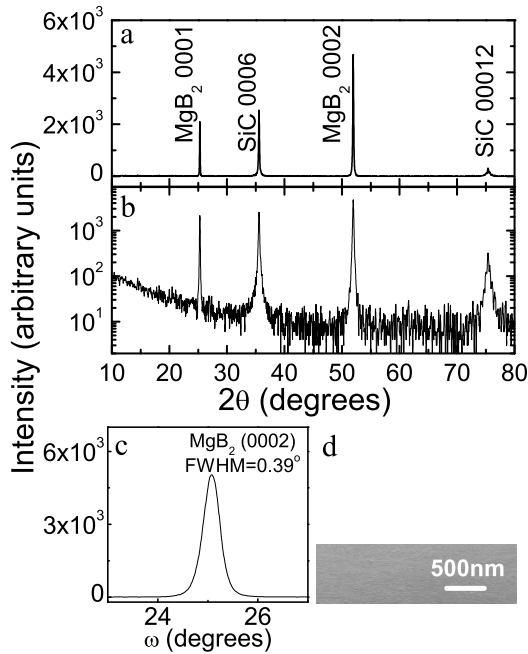}
\caption{(a) X-ray diffraction pattern of the MgB$_2$ film on a
(0001) 6$H$-SiC substrate in the $\theta$-2$\theta$ scan, which
shows only the $000l$ peaks of MgB$_2$ in addition to substrate
peaks, indicating a phase-pure $c$-axis-oriented MgB$_2$ film. (b)
The semilogarithmic plot of the $\theta$-2$\theta$ scan. (c) The
rocking curve of the 0002 MgB$_2$ peak, which shows the FWHM of
about 0.39$^\circ$. (d) The SEM image of the MgB$_2$ film, which
shows the smooth surface without obvious granularity.} \label{fig0}
\end{figure}

The high-quality MgB$_2$ thin films studied in this work were
prepared by the hybrid physical-chemical vapor deposition
technique\cite{film} on (0001) 6$H$-SiC substrates. All the films
had $c$-axis orientation with the thickness of about $100\;$nm.
Fig.~\ref{fig0} (a) shows the $\theta$-$2\theta$ scan of the MgB$_2$
film, and the sharp $(000l)$ peaks indicate the pure phase of the
$c$-axis orientation of MgB$_2$. In order to show the good
crystallinity of the film, we present in Fig.~\ref{fig0}(b) the same
data in a semilogarithmic scale which enlarges the data in the
region of small magnitude. It is clear that, besides the background
noise, we can only observe the diffraction peaks from MgB$_2$ and
the SiC substrate, i.e., there is no trace of the second phase in
the film. The $c$-axis lattice constant calculated from the MgB$_2$
peak positions was about 3.517$\;$\AA (bulk value\cite{MgB2found}:
3.524$\;$\AA). The $\phi$ scan (azimuthal scan) shown
elsewhere\cite{film} indicated well the sixfold hexagonal symmetry
of the MgB$_2$ film matching the substrate. The full width at half
maximum (FWHM) of the 0002 peak taken on the film in
$\theta$-$2\theta$ scan [MgB$_2$ 0002 peak in Fig.~\ref{fig0}(a)]
and $\omega$ scan [rocking curve, shown in Fig.~\ref{fig0}(c)] is
0.15$^\circ$ and 0.39$^\circ$, respectively. The scanning electron
microscopy (SEM) image in Fig.~\ref{fig0}(d) gave a rather smooth
top surface view without any observable grain boundaries, which
suggested that the film had a homogeneous quality. Ion etching was
used to pattern a four-lead bridge with the effective size of
$380\times20\;\mu\mathrm{m}^2$. The resistance measurements were
made in an Oxford cryogenic system Maglab-Exa-12 with magnetic field
up to 12$\;$T. Magnetic field was applied along the $c$ axis of the
film for all the measurements. The temperature stabilization was
better than $0.1\%$ and the resolution of the voltmeter was about
10$\;$nV. We have done all the measurements on several MgB$_2$
films, and the experimental data and scaling behaviors are similar;
so, in this paper, we present the data from one film.

\begin{figure}
\includegraphics[width=8cm]{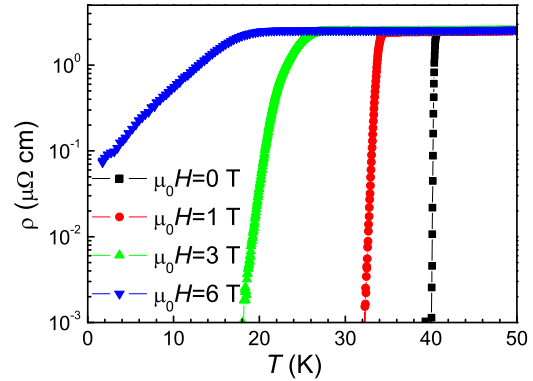}
\caption{Temperature dependence of resistive transitions for
$\mu_0H=0,\;1,\;3,$ and 6$\;\mathrm{T}$, with the current density
$j=500\;\mathrm{A/cm}^{2}$.} \label{fig1}
\end{figure}

In Fig.~\ref{fig1}, we present the resistive transitions ($R$-$T$
relations) of a MgB$_2$ thin film measured at various magnetic
fields in a semilogarithmic scale. The current density in the
measurement was about $500\;\mathrm{A/cm}^{2}$, much smaller than
the critical value for low temperatures, $10^6\;\mathrm{A/cm}^{2}\
$\cite{WenHHPRB2001}. It can be determined from Fig.~\ref{fig1} that
the sample had a superconducting transition temperature of
$T_\mathrm{c}=40.05\;$K, with a transition width of about $0.5\;$K.
Its normal state resistivity was about $2.45\;\mu\Omega\,$cm and the
residual resistance ratio [$\equiv
\rho(300\;\mathrm{K})/\rho(42\;\mathrm{K})$] was about $6.4$. The
$I$-$V$ curves were measured at various temperatures for each field,
and then we got the electric field ($E$) and the current density
($j$) according to the sample dimension. The current density was
swept from 5 to $10^5\;\mathrm{A/cm}^{2}$ during the $I$-$V$
measurements.

\section{Theoretical models}
In the mixed state of high-$T_\mathrm{c}$ superconductors with
randomly distributed pointlike pinning centers, a second-order phase
transition is predicted between VG state and vortex-liquid
state.\cite{Fisher_theory} The $I$-$V$ curves at different
temperatures near the VG transition temperature $T_\mathrm{g}$ can
be scaled onto two different branches\cite{Fisher_experiment} by the
scaling law
\begin{equation}
\frac{E}{j\left(T-T_\mathrm{g}\right)^{\nu(z+2-D)}}=f_{\pm}\left(\frac{j}{\left|T-T_\mathrm{g}\right|^{\nu(D-1)}}\right).
\end{equation}
The scaling parameter $z$ has the value of 4--7, and $\nu\approx$
1--2; $D$ denotes the dimension of the system with the value 3 for
3D and 2 for quasi-2D \cite{quasi2D}; $f_{+}$ and $f_{-}$ represent
the functions for two sets of the branches above and below
$T_\mathrm{g}$. Above $T_\mathrm{g}$, the linear resistivity is
given by
\begin{equation}
\rho_\mathrm{lin}=\left.\mathrm{d}E/\mathrm{d}j\right|_{j\rightarrow
0} \propto\left(T-T_\mathrm{g}\right)^{\nu(z+2-D)}.
\end{equation}

At $T_\mathrm{g}$, the electric field versus the current density
curve satisfies the relationship
\begin{equation}
E(j)|_{T=T_\mathrm{g}}\approx j^{(z+1)/(D-1)}.
\end{equation}

In 2D superconductors at $\mu_0H=0\;$T, a
Berezinskii-Kosterlitz-Thouless (BKT) transition was found at a
specific temperature $T_\mathrm{BKT}$.\cite{BKT} At
$T_\mathrm{BKT}$, $E\propto j^3$, which is a sign of the BKT
transition. A continuous change from the BKT transition at zero
field to a quasi-2D VG transition, and then to a true 2D VG
transition with $T_\mathrm{g}=0\;$K was found in TlBaCaCuO
film,\cite{Wen_TBCCO} which shows a field-induced crossover of
criticalities.

A 2D VG transition may exist in a true 2D system with
$T_\mathrm{g}=0\;$K, i.e., there is no zero-resistance state at any
finite temperatures. The $E$-$j$ curves can be scaled by\cite{2D_VG}
\begin{equation}
\frac{E}{j}\exp\left[\left( \frac{T_0}{T} \right) ^p \right]=g\left(
\frac{j}{T^{1+\nu_{2\mathrm{D}}}} \right),
\end{equation}
where $T_0$ is a characteristic temperature,
$\nu_{2\mathrm{D}}\approx2$, and $p\ge 1$, while $g$ is a scaling
function for all temperatures at a given magnetic field. The linear
resistance is given by
\begin{equation}
\rho_\mathrm{lin}\propto \exp[-(T_0/T)^p].
\end{equation}
This 2D scaling law can be achieved in the very thin
films\cite{2D_YBCO} or in highly anisotropic systems at high
magnetic fields.\cite{WenHHPRL97,WenHHPRL98}

\section{Experimental results and discussions}
\subsection{Quasi-two-dimension vortex-glass scaling in the low-field region ($\mu_0 H < $ 5 T)}
\begin{figure}
\includegraphics[width=8cm]{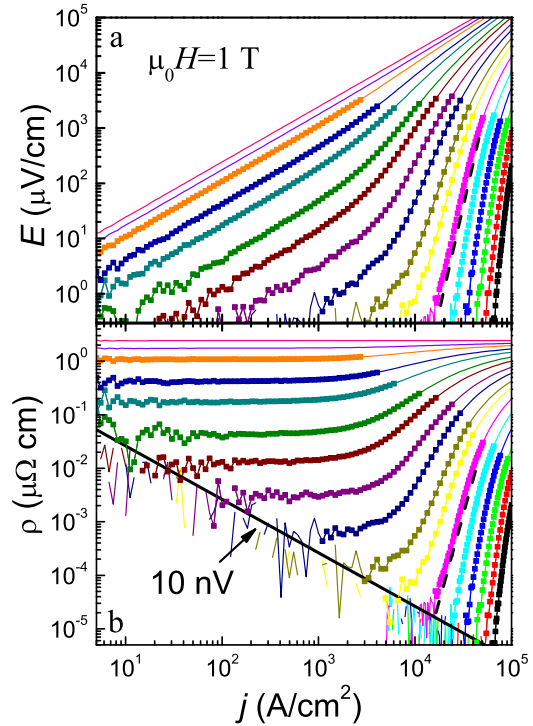}
\caption {(Color online) (a) $E$-$j$ characteristics measured at
fixed temperatures ranging from 30 to 36$\;$K for
$\mu_0H=1\;\mathrm{T}$. The increments are 0.30$\;$K in the range
from 30.00 to 31.20$\;$K, and 0.25$\;$K in the range from 31.50 to
34.00$\;$K respectively, and finally 35$\;$K on the top. The dashed
line shows the position of $T_\mathrm{g}$, and the symbols denote
the segments that scale well according to the quasi-2D VG theory.
The thin solid lines denote also the measured data, however, located
outside the scalable range. (b) $\rho$-$j$ curves corresponding to
the $E$-$j$ data in (a). The thick solid line in (b) denotes the
voltage resolution of 10$\;$nV.} \label{fig2}
\end{figure}

The $E$-$j$ characteristics have been measured at various magnetic
fields up to 12 T. In Fig.~\ref{fig2} we show the typical example at
$\mu_0H$ = 1 T for (a) $E$-$j$ curves and (b) the corresponding
$\rho$-$j$ curves in double-logarithmic scales. It is obvious that
when the temperature goes below some particular value (this is
actually the vortex-glass transition temperature $T_\mathrm{g}$
according to following discussions), the resistivity falls rapidly
with decreasing current density and finally reaches the
zero-resistance state which is the characteristic of the so-called
VG state. At the temperatures above $T_\mathrm{g}$, the resistivity
remains constant in small current limit. The current density of
$500\;\mathrm{A/cm}^{2}$ used in $\rho$-$T$ measurement shown in
Fig.~\ref{fig1} lies in this linear resistivity regime from about
$10^{-3}$ to $1\;\mu\Omega\,$cm. Consequently, these data sets
provide the basic information on scaling if the data are describable
by the VG theory.

\begin{figure}
\includegraphics[width=8cm]{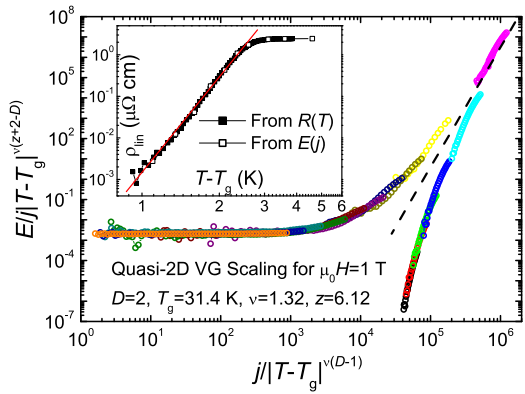}
\caption{(Color online) Quasi-2D VG scaling of the $E$-$j$ curves
measured at 1$\;$T. The inset shows a double-logarithmic plot of the
temperature dependence of the linear resistivity. The dashed line is
a guide for the eyes.} \label{fig3}
\end{figure}

The inset in Fig.~\ref{fig3} shows the data of the
$\rho_\mathrm{lin}$ versus $(T-T_\mathrm{g})$ and the fit to
Eq.~(2). The data are the same as those shown in Fig.~\ref{fig1} for
$\mu_0H=1\;$T, and the attempt $T_\mathrm{g}$ value is $31.4\;$K. In
this double-logarithmic plot, the slope of the linear fitting gives
just the exponent of $\nu(z+2-D)$, and the determined value is
$8.08\pm0.05$. In order to have reasonable values for $\nu$ and $z$,
the dimension parameter $D$ needs to be chosen as 2, i.e., the
investigated system has the property of quasi-2D, which is similar
to the situation found in BiSrCaCuO.\cite{quasi2D, zhang} This is
further supported by the VG scaling of the data at 1$\;$T. As shown
in the main frame of Fig.~\ref{fig3}, the scaling experimental
$E$-$j$ curves form two universal branches corresponding to the data
above and below $T_\mathrm{g}$ ($31.4\;$K) with $\nu=1.32$ and
$z=6.12$. At very large current density or a temperature near the
onset of superconducting transition, the free flux flow regime
dominates and, hence, the data do not scale. The symbols in the
figure denote the range of the data well described by the scaling
law.

\begin{figure}
\includegraphics[width=8cm]{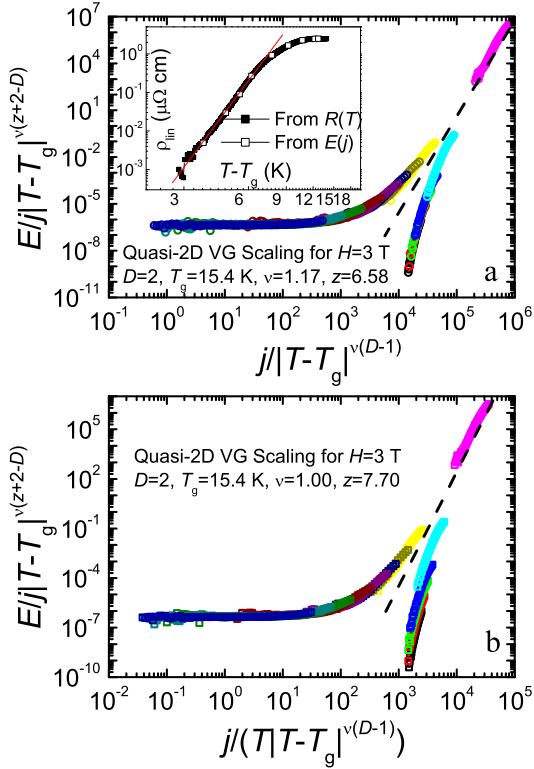}
\caption{(Color online)(a) Scaling curves of the $E$-$j$ data
measured in 3$\;$T based on the quasi-2D VG scaling theory. The
inset shows a log-log plot of the temperature dependence of the
linear resistivity. (b) VG scaling with another form of scaling
variable $j/(T\left|T-T_\mathrm{g}\right|^{\nu(D-1)})$ .}
\label{fig4}
\end{figure}

The situation at $\mu_0H=3\;$T is similar to that at $\mu_0H=1\;$T.
As shown in Fig.~\ref{fig4}(a), the determined parameters are
$T_\mathrm{g}=15.4\;$K, $\nu=1.17$, and $z=6.58$. Interestingly, the
previous work on MgB$_2$ film~\cite{MgB2IVfilm} indicated that the
3D VG scaling theory ($D=3$) is a better choice in describing the
$I$-$V$ characteristics in this system, though this experiment was
done at magnetic fields lower than 1$\;$T. Moreover, the $I$-$V$
curves were demonstrated to scale well by using the argument of
$j/(T\left|T-T_\mathrm{g}\right|^{2\nu})$. The same conclusions were
also drawn on the polycrystalline MgB$_2$ samples \cite{MgB2IVpoly}.
In order to clarify this issue, we also analyzed our data using the
form suggested in Ref.~8. As shown in Fig.~\ref{fig4}(b), such a
scaling with $j/(T\left|T-T_\mathrm{g}\right|^{\nu(D-1)})$ as the
scaling variable is worse than that with
$j/\left|T-T_\mathrm{g}\right|^{\nu(D-1)}$. Most importantly, the
dimension parameter $D$ is still required to be 2 instead of 3 as
proposed in Refs.~7 and 8. This confusion can be easily understood
in terms of the two-band superconductivity of MgB$_2$. As we know,
there are two types of bands contributing to the superconductivity
of MgB$_2$, namely, the 3D $\pi$ bands and the 2D $\sigma$ bands.
Therefore, the structure of the vortex matter must be affected by
both of them. Although the superconductivity of $\pi$ bands, induced
possibly by that of $\sigma$ bands, is much weaker, it provides a
large coherence length with 3D characteristics in the low-field
region. Therefore, the vortices in this system may be quasi-2D like
and, at the same time, they can possess large cores characterized by
the coherence length of the $\pi$ band superfluid. In this sense,
the quasi-2D scaling should be more appropriate than the 3D one.
However, when a higher disorder is induced in the system, especially
in the boron sites, the interband scattering gets stronger and the
anisotropy decreases, which may lead to a 3D vortex scaling. In this
case, a more rigid vortex line can be observed, especially at low
fields.\cite{JinH} The good quasi-2D scaling at 1 and 3$\;$T
demonstrated here suggests that the phase transition from VG to the
vortex liquid in MgB$_2$ resembles that in the high-$T_\mathrm{c}$
superconductors. Together with the data shown below, we can safely
conclude that a vortex glass state with zero linear resistivity can
be achieved in the low field region due to the presence of the
finite superfluid density from the $\pi$ bands. Regarding the VG
scaling\cite{Lobb} a principal requirement is a proper determination
of $T_\mathrm{g}$, namely the temperature with a straight $\log
E$-$\log j$ curve in the low dissipation part. The tolerance for
$T_\mathrm{g}$ variation is very small (about $\pm0.3\;$K). With an
inappropriately chosen $T_\mathrm{g}$, the scaling quality
dramatically deteriorates and, simultaneously, the values of $\nu$
and $z$ quickly deviate from those reported above and those proposed
by theory. This validates our analysis here.

\subsection{Anomalous vortex properties in high field region}

\begin{figure}
\includegraphics[width=8cm]{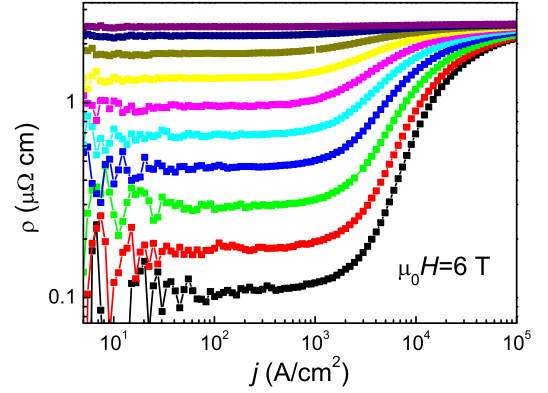}
\caption{(Color online) $\rho$-$j$ data at temperatures 1.7$\;$K and
4$\;$K to 20$\;$K with 2$\;$K-step, for $\mu_0H=6\;$T. Temperature
of the isotherms increases from bottom to top.} \label{fig5}
\end{figure}

\begin{figure}
\includegraphics[width=8cm]{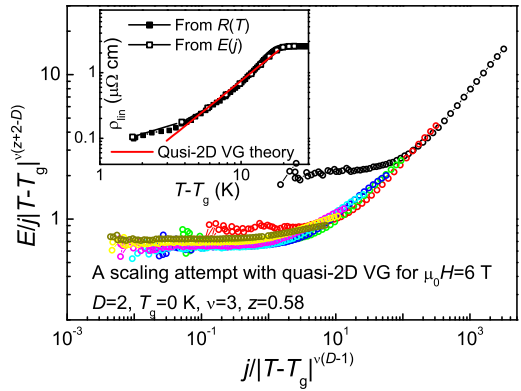}
\caption{(Color online) The scaling of the $E$-$j$ isotherms with
quasi-2D VG model for $\mu_0H=6\;$T. The inset shows the deviation
of the $\rho_\mathrm{lin}$ vs $T-T_\mathrm{g}$ ($T_\mathrm{g}=0$)
relation from the linearity in the double logarithmic scale.}
\label{fig6}
\end{figure}

\begin{figure}
\includegraphics[width=8cm]{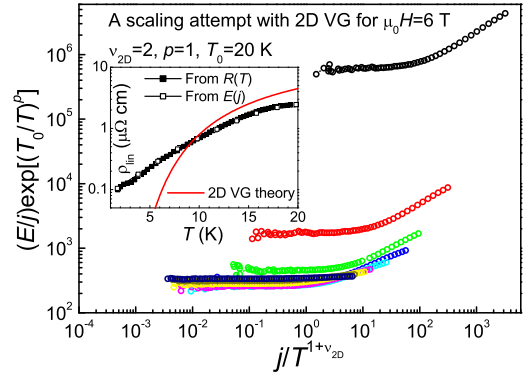}
\caption{(Color online) Attempted scaling of the data with 2D VG
model [Eq.~(4)] for $\mu_0H=6\;$T. The inset shows the nonlinearity
of the relationship between $\rho_\mathrm{lin}$ and temperature, the
solid line shows the theoretical curve of true 2D VG theory
[Eq.~(5)]. } \label{fig7}
\end{figure}

As shown in Fig.~\ref{fig1}, when the magnetic field reaches 6$\;$T,
no zero-resistance state can be observed down to the lowest
temperature, here 1.7$\;$K. Consequently, no VG transition exists
above 1.7$\;$K at this field, as shown in Fig.~\ref{fig5}. The shape
of the curve at $T=1.7\;$K suggests that the resistivity goes to a
finite value as the current density approaches zero.\cite{JiaY} As
shown in Fig.~\ref{fig6}, the $\rho_\mathrm{lin}$ versus
$T-T_\mathrm{g}$ seriously deviates from linearity for any possible
$T_\mathrm{g}$ value, indicating the inapplicability of Eq.~(2) in
the present case. Correspondingly, the quasi-2D scaling law fails
here. A natural explanation is that, with increasing field, the 3D
supercurrent from $\pi$ bands is seriously suppressed
\cite{gaps2,tunnelling} and the quasi-2D vortex structure transforms
into a 2D-like one dominated by the $\sigma$ band superfluid. In
Fig.~\ref{fig7}, we show our attempt to apply 2D VG scaling on the
data. Surprisingly, this attempt also failed, even though this model
has been successfully applied to the layered superconductors with
large anisotropy (or 2D property) such as Tl- and Bi-based
high-$T_\mathrm{c}$ thin films at high magnetic
fields.\cite{WenHHPRL97,WenHHPRL98}

The most reasonable explanation for this anomaly is that the
supercurrent contribution from the $\pi$ bands is much easier to
suppress by the magnetic field than that from the $\sigma$ bands,
since the gap in the $\pi$ bands is several times smaller than that
in the $\sigma$ band. We suggest that at high magnetic fields (above
5$\;$T), a different vortex matter state is formed, composed of
quasi-particles from the $\pi$ bands and vortices formed mainly by
the residual superfluid from the $\sigma$ band. The $\pi$-band
quasiparticles diminish the long range phase coherence of the
superconducting phase, which leads to a finite dissipation. Once the
long range superconducting phase coherence is destroyed by the
proliferation of a large amount of these $\pi$-band quasiparticles,
neither 3D nor quasi-2D VG scaling is applicable. Such a mixed state
is obviously difficult to be simply described by any known scaling
theory. Recently, scanning tunneling microscopy studies showed that
the quasiparticles of the $\pi$ bands disperse over all of the
superconductor, both within and outside the vortex
cores\cite{STMvortex}, which strongly supports our arguments. This
is the basis for the explanation of the nonvanishing vortex
dissipation at high magnetic fields in a zero temperature limit
found recently on MgB$_2$ thin films.\cite{JiaY}

\begin{figure}
\includegraphics[width=8cm]{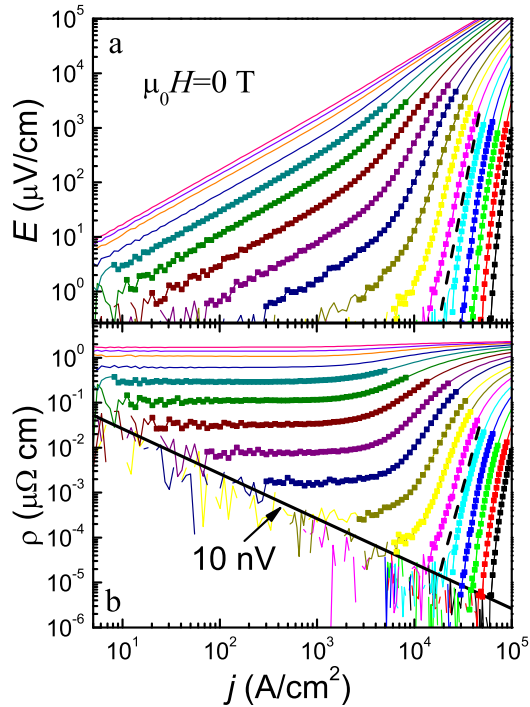}
\caption{(Color online) (a) $E$-$j$ data at various temperatures
from 39.7$\;$K to 40.5$\;$K with an interval of 0.05$\;$K for
$\mu_0H=0\;\mathrm{T}$, the symbols denote the region, where the
data are scaled (from 39.70$\;$K to 40.30$\;$K). Temperature of the
isotherms increases from bottom to top. The dashed line shows the
position of $T_\mathrm{g}$ and the symbols denote the segments,
which scale well according to the quasi-2D VG theory. The thin solid
lines are also the measured data lying outside the scalable range.
(b) $\rho$-$j$ curves corresponding to the $E$-$j$ data in (a). The
thick solid line in (b) denotes the voltage resolution of 10$\;$nV.}
\label{fig8}
\end{figure}

\begin{figure}
\includegraphics[width=9cm]{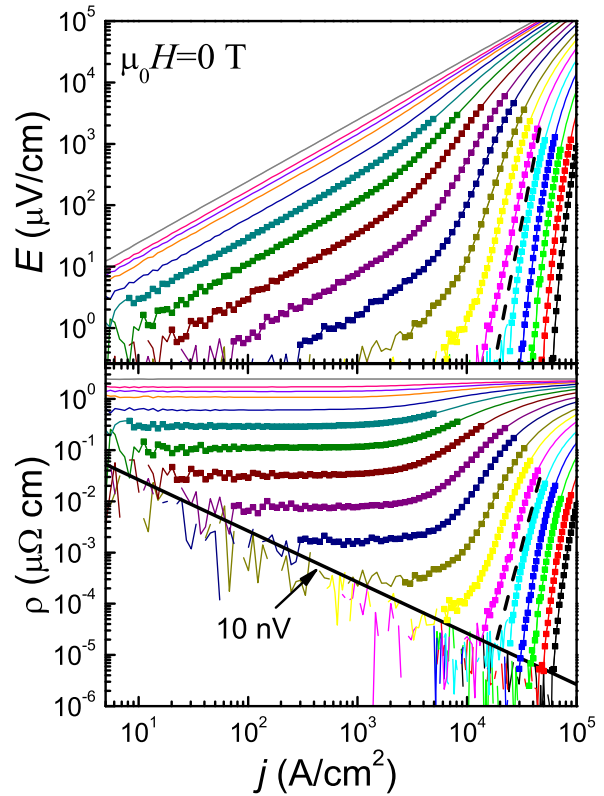}
\caption{(Color online) (a) Quasi-2D VG scaling of the data measured
at 0$\;$T. The inset indicates a good linearity of the temperature
dependence of the linear resistivity. (b) Quasi-2D VG scaling of the
data measured at 0.1$\;$T. The inset indicates a good linearity of
the temperature dependence of the linear resistivity.} \label{fig9}
\end{figure}

\subsection{Self-field effect at $\mu_0$H = 0 T}

For a 2D layered superconductor in zerofield, the above mentioned
BKT transition may exist and be reflected in the $I$-$V$
characteristics \cite{Wen_TBCCO}. In the present MgB$_2$ samples, we
have not found any evidence of this transition in low magnetic
fields which would be consistent with the quasi-2D (instead of 2D)
configuration of the vortex matter. Moreover, both the $E$-$j$
curves and the $\rho$-$j$ curves (as presented in Fig.~\ref{fig8})
are similar to the situation of $\mu_0H=1\;$T. Considering the
narrow transition width at zero-field, we did the measurement
carefully with an increment of 0.05$\;$K. Obviously, there is no
$E(j)$ curve which satisfies the $E\propto j^3$ dependence, as
expected by the BKT theory. Since the current can induce
self-generated vortices, it might be interesting to look at whether
the quasi-2D VG model applies here.

Similar in Sec.~IV~A, we present $\rho_\mathrm{lin}$ versus
$(T-T_\mathrm{g})$ in a double logarithmic plot. From this graph, we
determined the exponent in Eq.~2 (as shown by the inset of
Fig.~\ref{fig9}(a)). A good quasi-2D scaling was obtained with
parameters $T_\mathrm{g}=39.94\;$K, $\nu=1.12$, and $z=6.61$, as
presented in Fig.~\ref{fig9}(a). Using the parameters determined
here, one finds a self-consistency with the value of $\nu(z+2-D)$,
as determined in fitting the linear resistivity [Eq.~(2)]. Both the
temperature dependence of $\rho_\mathrm{lin}$ and the scaling curves
at $\mu_0 H=0\;$T are similar to the situation at small field $\mu_0
H=0.1\;$T [shown in Fig.~\ref{fig9}(b)] and $\mu_0 H=0.5\;$T (not
shown in this paper), except for the slight differences of the
scaling parameters. The scaling parameters including the ones at
$\mu_0H=0.5\;$T are listed in Table~\ref{table}. It was proven that
current and magnetic field exhibit analogous effects in suppressing
superconductivity and generating quasiparticles in conventional
superconductors.\cite{HIDOS} Similarly, the current-induced
self-field may lead to a similar effect in the vortex state as an
applied magnetic field. Nonetheless, the good agreement of this
simple scaling law with the zero-field data is interesting and worth
studying in detail. Moreover, the values of $\nu$ and $z$ for zero
field are very close to those for $\mu_0H=1$ and $3\;$T, indicating
a similar vortex dynamics in the whole low-field region.
\begin{table}
\caption{Quasi-2D VG scaling parameters at different fields.}
\begin{tabular*}{7cm}{@{\extracolsep{\fill}}cccc}
\hline \hline
$\mu_0H$ (T) & $T_\mathrm{g} (K)$    & $\nu$   &  $z$   \\
\hline
0.0   &     39.94   &   1.12    &   6.61  \\
0.1   &     39.28   &   1.30    &   6.08  \\
0.5   &     35.95   &   1.37    &   6.42  \\
1.0   &     31.4    &   1.32    &   6.12  \\
3.0   &     15.4    &   1.17    &   6.58  \\
\hline \hline
\end{tabular*}
\label{table}
\end{table}

\section{Summary}

We have measured $I$-$V$ curves on high-quality MgB$_2$ films at
various magnetic fields and temperatures. At magnetic fields below
$5\;$T including the zero field, the curves scaled well according to
the quasi-2D VG theory instead of the 3D model, in good agreement
with the multiband superconductivity of MgB$_2$ contributed from the
strong 2D $\sigma$ bands and weak 3D $\pi$ bands. At the fields
above $5\;$T, the curves did not scale according to any known VG
scaling laws, accompanied by the disappearance of a zero-resistance
state. Based on our result combined with recent tunneling
experiments, a different vortex state was suggested, namely, a state
where the vortices composed of the superfluid from the $\sigma$
bands move through the space filled with numerous quasiparticles
from $\pi$ bands.

\section{Acknowledgments}

This work is supported by the National Science Foundation of China,
the Ministry of Science and Technology of China (973 project:
2006CB601000 and 2006CB921802), and the Knowledge Innovation Project
of the Chinese Academy of Sciences (ITSNEM). The work at Penn State
is supported by NSF under Grants Nos. DMR-0306746 (X.X.X.),
DMR-0405502 (Q.L.), and DMR-0514592 (Z.K.L. and X.X.X.), and by ONR
under grant No. N00014-00-1-0294 (X.X.X.).

\end{document}